\documentclass[superscriptaddress,twocolumn,showpacs]{revtex4}
\usepackage[dvipdfmx]{graphicx}
\usepackage{xcolor}
\usepackage{dcolumn}
\usepackage{amsmath,mathtools}
\usepackage{dsfont}
\usepackage{epsfig,color}
\usepackage{soul}
\usepackage{amssymb}

\newcommand{\be}{\begin{equation}}
\newcommand{\ee}{\end{equation}}

\begin{document}

\title{Wave manipulation using a bistable chain with reversible impurities
}

\author{Hiromi Yasuda}
\affiliation{Department of Mechanical Engineering and Applied Mechanics, University of Pennsylvania, Philadelphia, PA 19104, USA}

\author{Efstathios G. Charalampidis}
\affiliation{Mathematics Department, California Polytechnic State University, San Luis Obispo, CA 93407-0403, USA}

\author{Prashant K. Purohit}
\affiliation{Department of Mechanical Engineering and Applied Mechanics, University of Pennsylvania, Philadelphia, PA 19104, USA}

\author{Panayotis G. Kevrekidis}
\affiliation{Department of Mathematics and Statistics, University of Massachusetts, Amherst, MA 01003-4515, USA}

\author{Jordan R. Raney \footnote{Corresponding author}}
\affiliation{Department of Mechanical Engineering and Applied Mechanics, University of Pennsylvania, Philadelphia, PA 19104, USA}

\pacs{}

\begin{abstract}
We systematically study linear and nonlinear wave propagation 
in a chain composed of piecewise-linear bistable springs. Such bistable systems
are ideal testbeds for supporting nonlinear wave dynamical features
including transition and (supersonic) solitary waves.
We show that bistable chains can support the propagation of 
subsonic wavepackets
which in turn can be trapped by a low-energy phase to 
induce energy localization.
The spatial distribution of these energy foci strongly affects the propagation of linear waves, typically causing scattering, but, in special cases, leading to a reflectionless mode analogous to the Ramsauer-Townsend (RT) effect. 
Further, we show that the propagation of nonlinear waves 
can spontaneously generate or remove additional foci, which act as effective
``impurities''.
This behavior serves as a novel mechanism for reversibly programming the dynamic response of bistable chains. 

\end{abstract}

\maketitle

\section{Introduction}
Multistable mechanical metamaterials have attracted increasing
attention from various research communities in physics, engineering,
and materials science, both because of their unique static
and dynamic behavior~\cite{Shan2015,Rafsanjani2015,Bertoldi2017}.
Indeed, multistability has been leveraged to achieve controlled state changes in engineering applications such as 
reconfigurable robots~\cite{Hawkes2010a,Chen2018}, deployable 
space structures~\cite{Seffen1999}, 
medical devices~\cite{Kuribayashi2006}, and many others. 

Even simple discrete systems composed of bistable elements (i.e., bistable lattices) are capable of rich and varied nonlinear dynamics.
For example, by tailoring the energy landscape of a bistable element, a bistable lattice can allow 
the propagation of transition fronts~\cite{Nadkarni2014,Nadkarni2016,Raney2016,Jin2020,Yasuda2020} 
where the propagating directions and speed can be manipulated.
In addition, recent studies have shown that a bistable chain can support the propagation of solitary waves~\cite{Katz2018,Katz2019,Vainchtein2020}.

Here, we study the \textit{in situ} manipulation of the dynamic
response of a bistable chain in the broader context of linear/nonlinear wave dynamics. 
In particular, we showcase a mechanism for controllably inserting reversible 
effective ``impurities'' in a bistable chain, arising 
from the trapping of propagating wavepackets.
We numerically investigate the amplitude-dependent behavior of a bistable chain that can support 
{not only supersonic solitary waves but, depending on the impact speed, also subsonically propagating
wavepackets.}
We find that the propagation of the latter is brought to a 
halt in the bulk of the chain, leading to the formation of effective 
impurities (albeit within a homogeneous medium) that may play a central role in 
subsequent dynamics.
More specifically, we demonstrate that the presence of such energy foci affects the scattering of linear waves. 
Interestingly, if two such effective 
impurities are placed next to each other and the system is driven at a specific frequency that leads to the propagation of a linear wave of a particular
wavenumber, complete transmission without reflection 
 can occur in a bistable lattice~\cite{Martinez2016}. 
 This is reminiscent of the  Ramsauer-Townsend (RT) effect, which is well-known in quantum mechanics~\cite{sakurai}, 
 the simplest instance of which involves the 
 perfect transmission at select energies 
 in one-dimensional scattering from a square well.
Multiple impurities can be created via collisions of a 
propagating wavepacket with an impurity.
Note that hereafter we will use the term impurity in this loose sense, implying 
the presence of a standing wave, even though no spatial inhomogeneity is present
in the lattice.

Our presentation is organized as follows: Section~\ref{Sec2} discusses 
the model setup, explains the dynamic response of a 1D bistable chain under impact
excitation, and 
identifies the formation of impurities due to the trapping of a subsonic wavepacket.
In Section~\ref{sec3}, 
we investigate the scattering behavior between plane waves and 
single impurities.
Section~\ref{Sec4} explores scattering for the case with multiple
impurities. 
Finally, Section~\ref{conclusion} summarizes our findings and presents 
directions for future studies.

\begin{figure}[htbp]
    \centerline{ \includegraphics[width=0.5\textwidth]{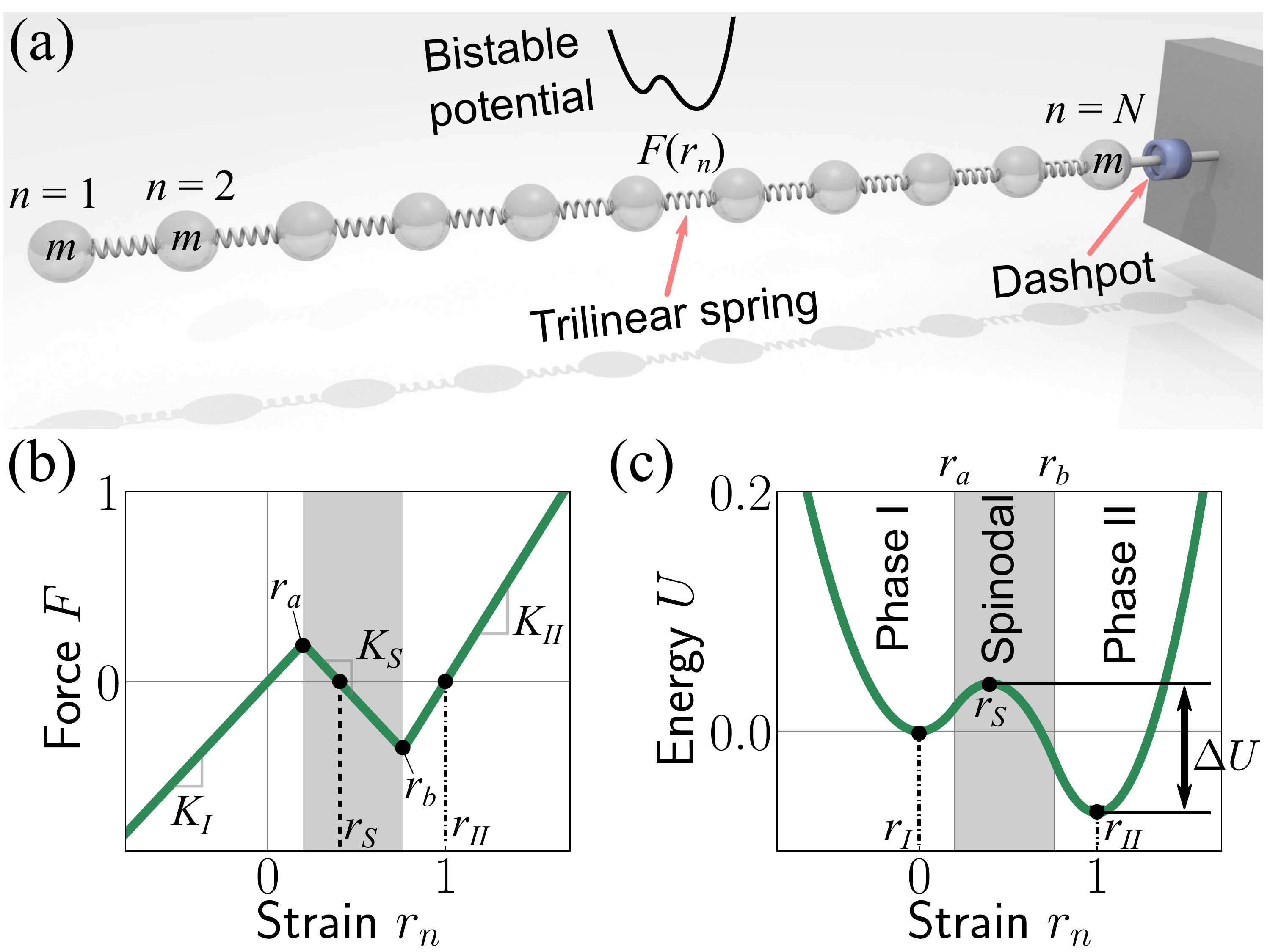}}
    \caption{(a) Schematic illustration of the system composed of bistable springs. 
    The impact is applied to the left-most unit ($n=1$) by setting an initial velocity. 
    The right-most mass ($n=N$) is connected to a dashpot. (b)~The relationship between force and strain 
    of a bistable spring is expressed by a trilinear function [cf. Eq.~\eqref{eq:trilinear}]. The stiffness parameters are $(K_{I}, K_S, K_{II})=(1, -1, 1.5)$ and the strain parameters are $(r_S, r_{II})=(0.4, 1)$. (c)~The bistable energy landscape obtained from the trilinear force-strain relationship of panel (b).
    }
    \label{fig:piece_wise}
\end{figure}

\section{Bistable chains}\label{Sec2}
\begin{figure*}[htbp]
\centering
\includegraphics[width=1.0\linewidth]{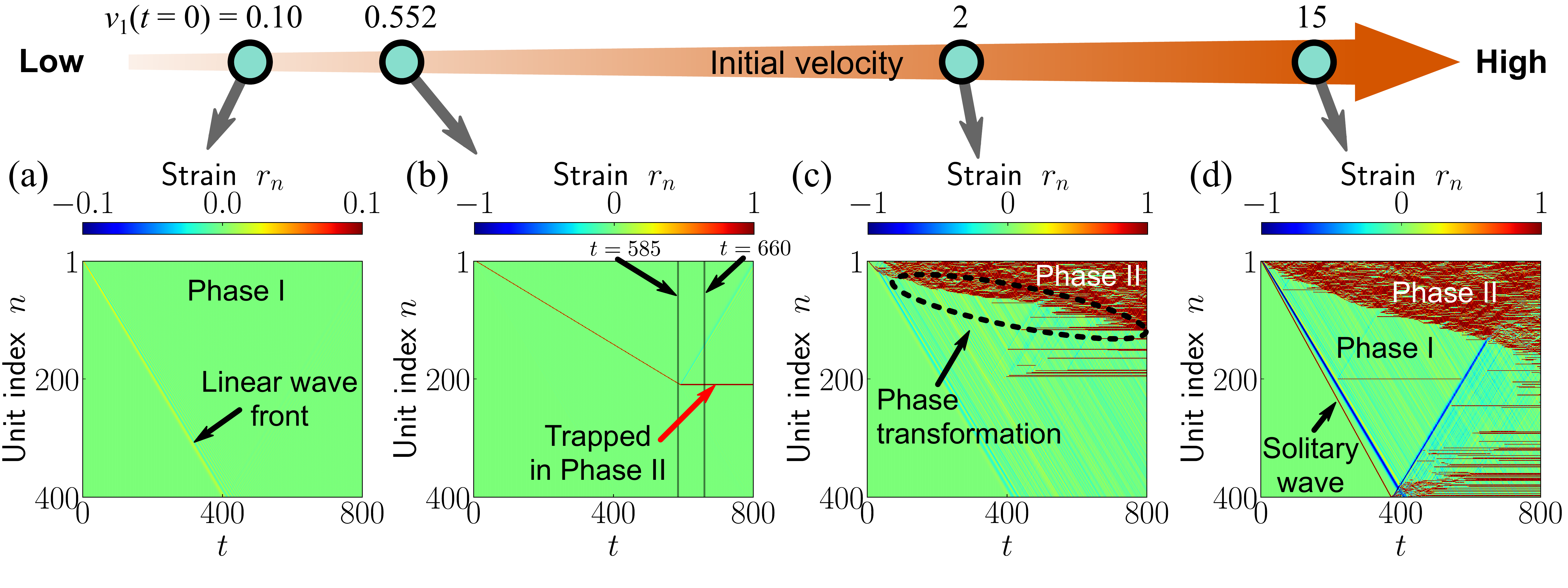}
\vspace{-25pt}
\caption{
Amplitude-dependent behavior for a bistable chain composed of strain-hardening springs ($K_{II}/K_{I}=1.5$). All unit cells are initially in Phase I. We apply 
an impact velocity to the first unit cell ($n=1$) by setting 
$v_1(t=0)$. 
We consider four different values of the impact velocities: 
(a)~$v_1(t=0)=0.1$, (b)~$0.552$, (c)~$2$, and (d)~$15$. 
The color indicates the strain 
 $r_n$ as a function of the unit index $n$ and time $t$. 
}
\label{fig:AmpDependent}
\end{figure*}



We  begin our discussion by examining the dynamic response of a 1D lattice 
under an elastic impact excitation 
caused by a striker 
(see Fig.~\ref{fig:piece_wise}(a)).
We define $u_n$ as the relative displacement of the $n$th 
particle measured from its equilibrium point. 
We define the 
strain of the $n$th particle as $r_n\coloneqq u_n-u_{n+1}$.
Motivated by the reconfigurability of the bistable lattice
we consider the trilinear spring element (see Fig.~\ref{fig:piece_wise}(b)) 
defined by
\begin{equation}
	F(r) = \left\{
		\begin{array}{lcr}
		{{K}_{I}}r &r \leq r_a\\
		K_S \left( r - r_S \right) &r_a \leq r \leq r_b\\
		K_{II} \left( r - r_{II} \right) &r_b \leq r
		\end{array}
	\right. ,
\label{eq:trilinear}
\end{equation}
where $K_{I}$, $K_{S}$, and $K_{II}$ are the stiffness parameters.
Therefore, Eq.~\eqref{eq:trilinear} is 
a double-well
potential function as 
shown in Fig.~\ref{fig:piece_wise}(c). 
In this study, we use the following set of numerical values for the 
parameters: 
$(K_{I}, K_S, K_{II})=(1, -1, 1.5)$, and 
$(r_S, r_{II})=(0.4, 1)$. By using these values, 
we can determine the other strain parameters $(r_a, r_b)$.
Here, we define the ``Phase~I'' and ``Phase II'' regimes 
corresponding to the stable states around $r_I=0$ and 
$r_{II}=1$, respectively (see also Fig.~\ref{fig:piece_wise}(c)). 
These two regimes are connected by the spinodal region corresponding to the negative stiffness range (see the shaded gray area in Fig.~\ref{fig:piece_wise}(b)).

Based on this trilinear spring element, the dynamic response of our bistable chain 
is analyzed by using the following equation of motion:
\begin{equation} \label{eq:EqMo_Mono}
{m}{{\ddot{u}}_{n}}=F\left( {{u}_{n-1}}-{{u}_{n}} \right)%
-F\left( {{u}_{n}}-{{u}_{n+1}} \right),
\end{equation}
where $m$ is the mass of the particle whose value is fixed 
to $m=1$, and $n=1,\dots, N$ with $N$ being the number 
of particles in the chain.
It should be noted that we can calculate two sound speeds: $c_{I}=\sqrt{K_{I}/m}=1$
and $c_{II}=\sqrt{K_{II}/m}=\sqrt{1.5}$ corresponding to Phase I and Phase II,
respectively. 

\subsection{Amplitude-dependent behavior}\label{sec:amp}

We analyze strain profiles of propagating waves in 
a chain consisting of $N=400$ particles under impact.
We numerically simulate the effect of
impact by setting the velocity of the first particle to a non-zero value, that is, we set $v_{1}(t=0)\neq 0$. All other initial conditions are set to zero
(i.e., positions
$u_{n}(t=0)$ and velocities $v_{n}(t=0)$).
Then, we advance Eq.~\eqref{eq:EqMo_Mono} forward in time by 
using a standard fourth-order Runge-Kutta method.
To minimize the effect of reflected waves at the other 
end of the chain ($n=400$), the last particle is connected 
to a dashpot in which the damping force is linearly proportional 
to the velocity of the last particle ($F_{\nu}=\nu v_{N}$ with
$\nu=1$). 
The right end of the chain is attached to a rigid 
wall, i.e., $u_{401}(t)\equiv v_{401}(t)=0\,\, \forall t\geq 0$.

We now discuss Fig.~\ref{fig:AmpDependent} which 
shows the space-time contour plots of the strain 
variable for four different values of impact velocities:
$v_1(t=0)=0.1$, $0.552$, $2$, and $15$. In particular, if
$v_1(t=0)=0.1$, we only observe small-amplitude wave propagation
and the chain configuration remains in Phase~I (see
Fig.~\ref{fig:AmpDependent}(a)). If we increase 
the impact 
velocity to $v_1(t=0)=0.552$,
we observe the formation
of a propagating localized wavepacket with amplitude (in the strain variable) greater than $r_{S}=0.4$ 
(see Fig.~\ref{fig:AmpDependent}(b)). 
This finding indicates that spring elements undergo a transition 
from Phase~I to Phase~II and then 
back to Phase~I as 
the wavepacket propagates in the chain.
It is noteworthy that the speed of this excitation
($V_{tw}=0.358$)
is slower than that of the
small-amplitude wave front (see Fig.~\ref{fig:AmpDependent}(a-b)). 
That is, this is a subsonic 
propagating wavepacket.
Interestingly, 
this 
structure is brought to
a halt before it reaches the end of the chain, 
and two adjacent spatial nodes
are trapped in Phase~II (see the red arrow in Fig.~\ref{fig:AmpDependent}(b)),
forming the effective impurity discussed above.
Note that we observe this trapping
behavior for the hardening case (i.e., $K_{II}/K_{I}=1.5$), 
however, the chain with softening springs (e.g., $K_{II}/K_{I}=0.5$)
does not show the formation of impurities or of propagating excitations
(see Appendix A for details). 
Although propagation
of supersonic solitary waves in a nonintegrable Fermi-Pasta-Ulam
chain~\cite{Friesecke_1999,Truskinovsky2014}, 
specifically a bistable
chain~\cite{Katz2018,Katz2019,Vainchtein2020}, has been reported previously,
the formation of impurities arising from the 
spontaneous trapping behavior we observe in the present work has been unexplored, to the best of our knowledge.

If $v_1(t=0)$ is further increased to $2$, the chain exhibits phase
transformation of multiple units 
which propagate from the left end of the chain to the other end as
shown in Fig.~\ref{fig:AmpDependent}(c). 
Finally, when we apply an
extremely large amplitude input to the system ($v_1(t=0)=15$),
we find both propagation of a solitary wave ($V_{tw}=1.076$) and a phase transformation front (see Fig.~\ref{fig:AmpDependent}(d)).
This solitary wave propagates faster than the (linear) sonic wave
speed (i.e., it is a supersonic solitary wave).
Therefore, we find that this bistable chain can support coherent structures 
of both subsonic and supersonic nature, depending on the magnitude of the impact velocity. 
We will discuss the trapping of subsonically propagating 
wavepackets in the next section.

\subsection{Trapping behavior}\label{sec:trapping}
\begin{figure*}[htbp]
    \centerline{ \includegraphics[width=1.\textwidth]{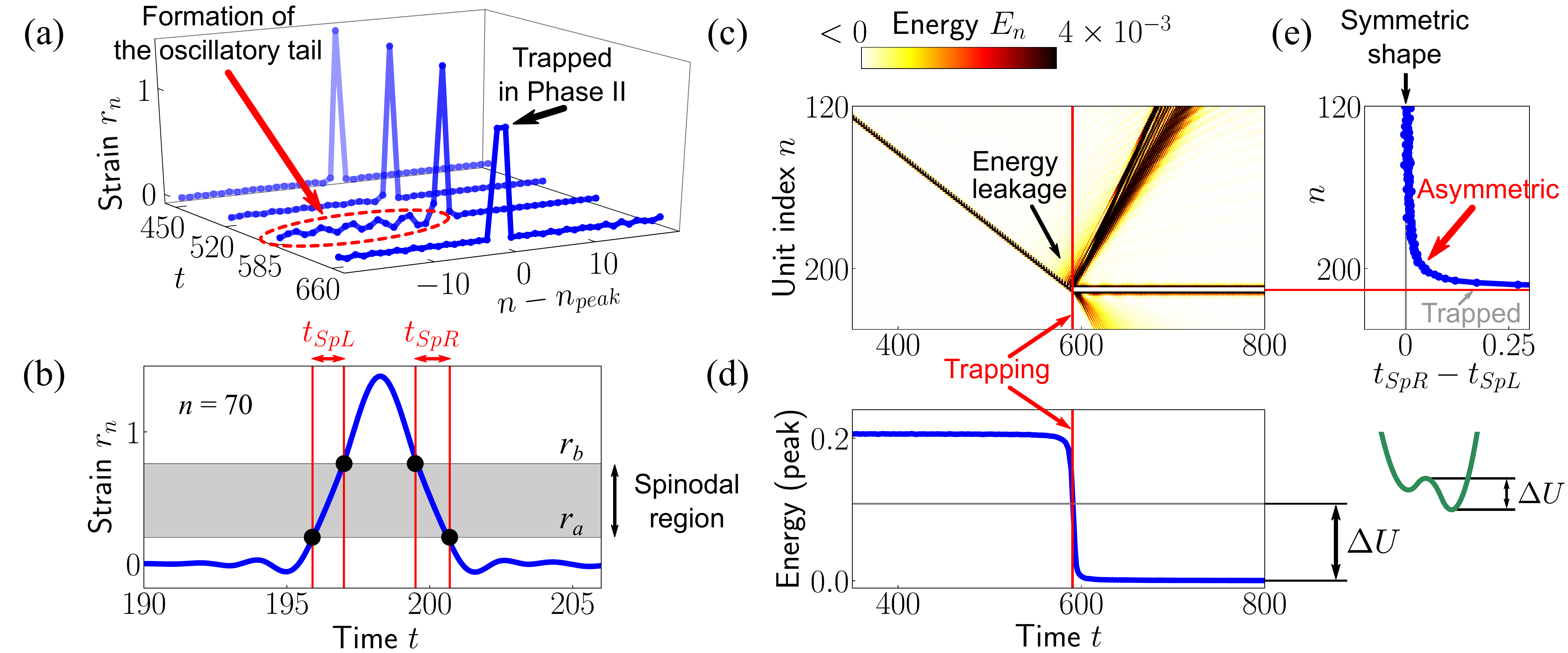}}
    \caption{Trapping behavior. (a) The spatial distribution of the subsonic wavepacket generated by the impact velocity $v_1(t=0)=0.552$ is shown at different instants of time $t$. 
    The waveforms for $t=585$ and $660$ correspond to the black vertical lines in Fig.~\ref{fig:AmpDependent}(b), respectively. Each strain profile is shifted in the space domain so that the peak of the strain profile is located at $n=0$ to ease visualization. (b) The strain change of $n=70$ is shown as a function of time $t$. The shaded gray 
    area indicates the spinodal regime (see also Fig.~\ref{fig:piece_wise}(c)). To quantify the wave shape, we measure time widths in which the strain is in the spinodal region, $t_{SpL}$ and $t_{SpR}$, bounded by the red vertical lines. (c) Surface plot of the energy $E_n(t)=\frac{1}{2}m \dot u_n^2(t) + U_n(t)$ where $U_n$ is the bistable potential energy function of 
    Fig.~\ref{fig:piece_wise}(c). (d) 
    The peak energy of the propagating wavepacket is shown as a function of $t$. $\Delta U$ is the energy barrier measured from the energy minimum in Phase II (see Fig.~\ref{fig:piece_wise}(c)). The red vertical line indicates the trapping point at which the peak energy becomes lower than that of the energy barrier $\Delta U$. (e) We calculate the time difference $t_{SpR}-t_{SpL}$, which indicates the symmetric (if $t_{SpR}-t_{SpL}=0$) or asymmetric waveform (otherwise). 
    }
    \label{fig:trapping}
\end{figure*}

To understand this trapping behavior thoroughly, we examine the time evolution of the subsonic propagating wavepacket by extracting the strain waveforms at different time frames, especially right before the trapping occurs.
In particular, 
Fig.~\ref{fig:trapping}(a) shows the strain profiles for four different time frames: $t=450$,
$520$, $585$, and $660$, (the shapes at $t=585$ and $660$ correspond to the black vertical lines in Fig.~\ref{fig:AmpDependent}(b)). 
Note that the wave profiles are shifted in the spatial domain so that the peak of the wave is located at the center of the domain to ease
visualization.
The subsonic localized pattern propagates without any noticeable 
distortions initially (see also Fig.~\ref{fig:trapping}(b) for the strain change of $n=70$ as a function of time $t$), 
although as time passes, the profile features an oscillatory tail as can be
discerned at $t=585$ (just before the wave stops) 
which is denoted by the red dashed ellipse 
in Fig.~\ref{fig:trapping}(a). 
This tail pattern of the wavepacket stems from its resonance
with the linear modes  associated with the (zero) background.

Indeed, as the wave propagates 
the profile loses energy 
in the form of emitted radiation.
This is investigated in 
panels (c) and (d) of Fig.~\ref{fig:trapping}, which depict the spatio-temporal evolution
of the total energy of the system and the peak of the energy
as a function of time, respectively. 
As the wave propagates
within the chain, the peak energy remains nearly constant, until an abrupt drop in the form of radiation immediately prior to trapping (i.e., when the total energy becomes lower than $\Delta U$, the energy barrier 
needed to be overcome to return to Phase~I). 
In addition to these energy considerations, we analyze the change of the wave shape by considering the strain in the spinodal region, which is associated with negative stiffness (Fig.~\ref{fig:piece_wise}(b)).
To characterize the wave shape, we measure two time widths, $t_{SpL}$ and $t_{SpR}$, in which the strain is in the spinodal region (Fig.~\ref{fig:trapping}(b)).
We plot the time width difference $t_{SpR}-t_{SpL}$ as shown in Fig.~\ref{fig:trapping}(e) 
in which a positive value of $t_{SpR}-t_{SpL}$ indicates that the strain stays longer in the spinodal region when the unit goes back to the initial state from Phase II.
The value of $t_{SpR}-t_{SpL}$ is nearly zero initially (i.e., symmetric wave shape). 
{However, this value increases significantly, most notably right before trapping. This is indicative of rapid growth of distortion.}
There are two competing processes: 1) potential energy release 
(and conversion to kinetic) due to the transition from Phase~I to Phase~II, and 2) potential energy 
re-balancing from the wavepacket due to its jumping over the energy barrier (and also the spinodal region). 
The propagation of the subsonic wavepacket shows that these two competing processes are not in the perfect balance
needed for genuine traveling. Rather, the energy 
lost in the form of leakage of radiation 
wavepackets dominates the energy 
release, as is evidenced by monitoring the spinodal region.
As a result, the wavepacket eventually gets stuck in Phase II, nucleating a new stable phase due to the metastability (cf. Refs.~\cite{Balk2001,Ngan2002}).
These act effectively as impurities with respect to wave propagation, as described in the next section.

\section{Scattering between plane waves and impurities}\label{sec3}
Based on the 
impurity formation described above, we now turn our focus to 
the scattering of a plane wave by such impurities in the system. 
We analyze the propagation of linear waves by considering the
following linearized equation of motion posed on an infinite lattice:
\begin{equation} \label{eq:LinearEq}
	m{{\ddot{u}}_{n}}={{K}_{n-1}}{{u}_{n-1}}+{{K}_{n}}{{u}_{n+1}}-\left( {{K}_{n-1}}+{{K}_{n}} \right){{u}_{n}},
\end{equation}
where $K_n$ is $K_{II}$ ($K_{I}$) for impurities (rest of the
particles). We can thus determine the normal modes of the system
by introducing the plane wave ansatz $u_n= e^{i\left(kn-\omega t\right)}$ 
where $k$ and $\omega$ are the wave number and  frequency, respectively 
(with $i=\sqrt{-1}$).
Without impurities, we can obtain the customary dispersion relation: 
\begin{equation}
	\omega = 2\pi f =\sqrt{\frac{2K_I}{m}\left[1-\cos(k)\right]} .
\end{equation}	

We systematically study scattering of plane waves by introducing a
single impurity (i.e., only a single spring element is in Phase~II) 
or double impurity (two adjacent elements in Phase~II) into a host chain in
which all elements are initially in Phase~I.
Figures~\ref{fig:analytics}(a) and \ref{fig:analytics}(b) show the
schematic illustrations for a chain with a single impurity
($K_{n=0}=K_{II}$) and double impurity ($K_{n=0}=K_{n=1}=K_{II}$).

\subsection{Theoretical analysis}\label{sec:theory}
\begin{figure*}[htbp]
    \centerline{ \includegraphics[width=1.0\textwidth]{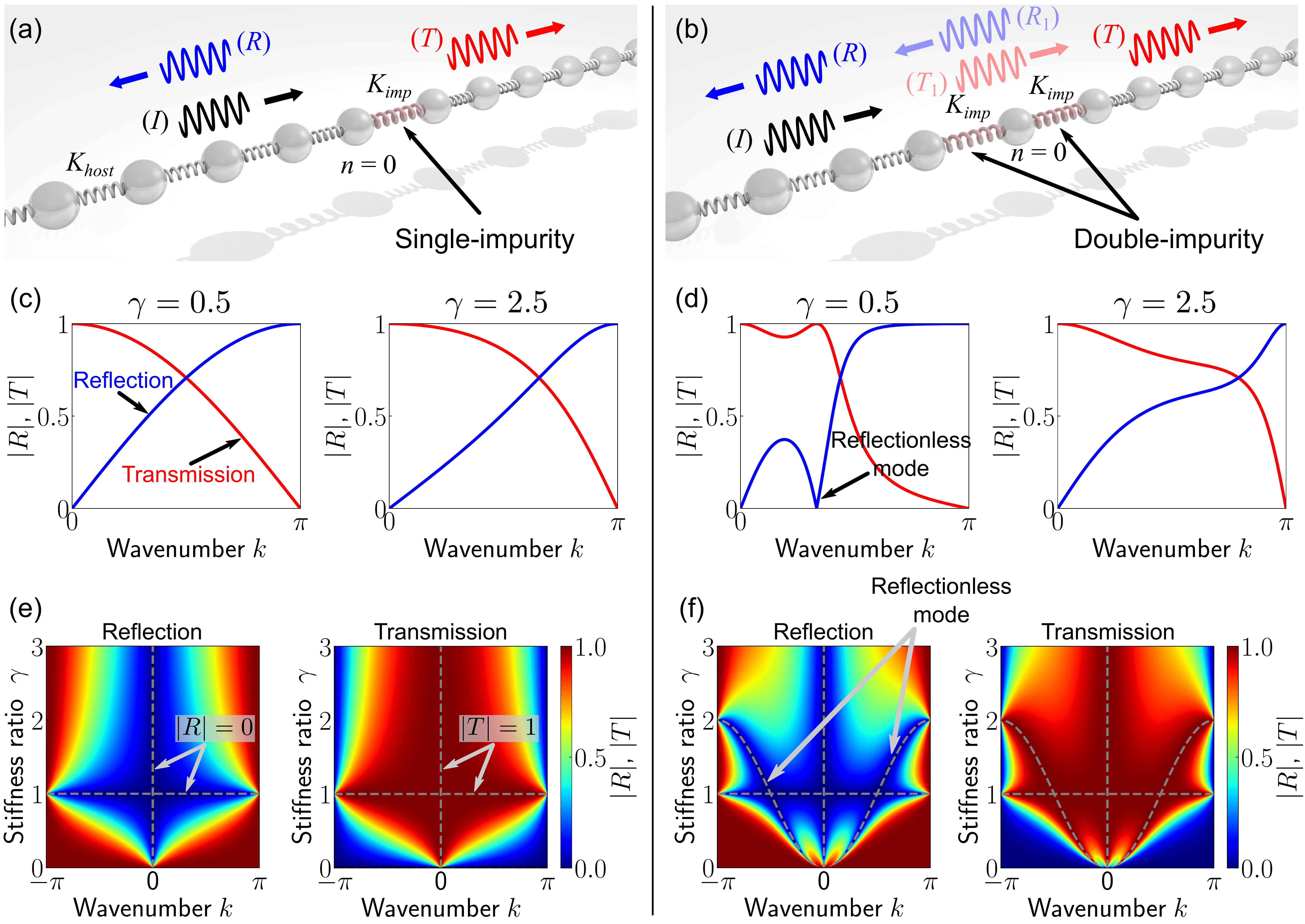}}
    \caption{(a-b) Schematic illustrations for a chain with a (a) single and (b) double impurity.  Analytical transmission and reflection coefficients for (c) single- and (d) double-impurity cases. The two different stiffness ratio cases (left) $\gamma=0.5$ and (right) $\gamma=2.5$, are shown. The reflection and transmission coefficients for (e) single- and (f) double-impurity cases are plotted as a function of stiffness ratio ($\gamma$) and wavenumber ($k$). The gray dashed lines indicate $|R|=0$ (or $|T|=1$. (e-f)
    }
    \label{fig:analytics}
\end{figure*}
To examine the effect of impurities on the scattering problem, we employ the following 
ansatz:
\begin{equation}
	u_n = \left\{
		\begin{array}{lcr}
		e^{i(kn-\omega t)}+Re^{-i(kn+\omega t)}\,,&\quad\text{if}
		&n\leq0\\
		Te^{i(kn-\omega t)}\,,&\quad\text{if}&n>0
		\end{array}
	\right.\,,
\label{eq:ansatz}
\end{equation}
composed of incident (denoted by $I$ in Fig.~\ref{fig:analytics}(a)),
reflected ($R$), and transmitted waves ($T$). This way,  
$|R|^2$ and $|T|^2$ are the reflection and transmission coefficients, respectively.
By following the same procedure used for the granular chain with Hertzian interactions~\cite{Martinez2016}, we plug 
Eq.~\eqref{eq:ansatz} into Eq.~\eqref{eq:LinearEq}, 
and obtain the linear system of equations: 
%
\begin{equation} \label{eq:matrixform}
	\mathbf{Ax}=\mathbf{b},
\end{equation}
with
%
\begin{subequations}
\begin{align}
	\mathbf{A} &= \left[ \begin{matrix}
   \gamma {{e}^{ik}} & \frac{ {{\omega }^{2}}m}{K_{I}}+{{e}^{ik}}-1-\gamma  \\
   \left( 1+ \gamma \right){{e}^{ik}}-{{e}^{2ik}}-\frac{ {{\omega }^{2}}m}{K_{I}}{{e}^{ik}} & -\gamma  \\
\end{matrix} \right],\\
	\mathbf{x} &= {{\left[ \begin{matrix}
   T, & R  \\
\end{matrix} \right]}^{T}},\\
	\mathbf{b} &= {{\left[ \begin{matrix}
   -\frac{ {{\omega }^{2}}m}{K_{I}}+1+\gamma-{{e}^{-ik}}, & \gamma  \\
\end{matrix} \right]}^{T}},
\end{align}
\end{subequations}
%
for the single-impurity case. Here, we define the stiffness 
ratio as $\gamma = K_{II}/K_{I}$. Solving Eq.~\eqref{eq:matrixform}
yields respectively the reflection and transmission coefficients:
\begin{subequations}
\begin{align}
	{{\left| R \right|}^{2}} &= {{\left| \frac{\left( 1-\gamma  \right)\left( -1+{{e}^{ik}} \right){{e}^{ik}}}{\left( -1+{{e}^{ik}} \right)-2\gamma {{e}^{ik}}} \right|}^{2}}, \\
	{{\left| T \right|}^{2}} &= {{\left| \frac{\gamma \left( 1+{{e}^{ik}} \right)}{\left( 1-{{e}^{ik}} \right)+2\gamma {{e}^{ik}}} \right|}^{2}}.
\label{eq:single}
\end{align}
\end{subequations}

For a chain with a double impurity, we use 
the following ansatz instead:
\begin{equation}
	u_n = \left\{
		\begin{array}{lcr}
		e^{i(kn-\omega t)}+Re^{-i(kn+\omega t)}&\quad\text{if}
		&n<0\\
		{{T}_{1}}{{e}^{i\left( kn-\omega t \right)}}+{{R}_{1}}{{e}^{-i\left( kn+\omega t \right)}}&\quad\text{if} &n=0,1\\
		Te^{i(kn-\omega t)}&\quad\text{if}&n>1
		\end{array}
	\right. .
\label{ansatz2}
\end{equation}
%
Here, $|R_1|^2$ and $|T_1|^2$ 
correspond to the reflection and transmission coefficients 
inside the double-impurity region (see Fig.~\ref{fig:analytics}(b)).
Then, following the same procedure as above, we arrive
at 
\begin{widetext}
\begin{subequations}
\begin{align}
	\mathbf{A} &= \left[ \begin{matrix}
   0 & {{e}^{2ik}}-\left( 1 + \gamma - \frac{ {{\omega }^{2}}m}{K_{I}} \right){{e}^{ik}} & \gamma & \gamma  \\
   0 & \gamma {{e}^{ik}} & \gamma {{e}^{ik}}-2\gamma+ \frac{ {{\omega }^{2}}m}{K_{I}} & \gamma {{e}^{-ik}}-2\gamma+\frac{ {{\omega }^{2}}m}{K_{I}}  \\
   {{e}^{2ik}} & 0 & \gamma-\left( \gamma+ 1 - \frac{ {{\omega }^{2}}m}{K_{I}} \right){{e}^{ik}} & \gamma-\left( \gamma+1-\frac{ {{\omega }^{2}}m}{K_{I}} \right){{e}^{-ik}}  \\
   \left( \frac{ {{\omega }^{2}}m}{K_{I}} -2 \right){{e}^{2ik}}+{{e}^{3ik}}& 0 & {{e}^{ik}} & {{e}^{-ik}}  \\
\end{matrix} \right],\\
	\mathbf{x} &= {{\left[ \begin{matrix}
   T, & R, & {{T}_{1}}, & {{R}_{1}}  \\
\end{matrix} \right]}^{T}},\\
	\mathbf{b} &= {{\left[ \begin{matrix}
   \left( -\frac{ {{\omega }^{2}}m}{K_{I}}+1+\gamma \right){{e}^{-ik}}-{{e}^{-2ik}}, & -\gamma {{e}^{-ik}}, & 0, & 0  \\
\end{matrix} \right]}^{T}}.
\end{align}
\end{subequations}
\end{widetext}
In this way, we have 
%
\begin{widetext}
\begin{subequations}
\begin{align}
	{{\left| R \right|}^{2}} &= {{\left| \frac{\left( 1-\gamma  \right)\left( -1+{{e}^{ik}} \right)\left\{ {{\left( -1+{{e}^{ik}} \right)}^{2}}+2\gamma {{e}^{ik}} \right\}}{-{{e}^{ik}}+\left( 3-4\gamma  \right){{e}^{2ik}}-3{{\left( 1-\gamma  \right)}^{2}}{{e}^{3ik}}+{{\left( 1-\gamma  \right)}^{2}}{{e}^{4ik}}} \right|}^{2}},\\
	{{\left| T \right|}^{2}} &= {{\left| \frac{{{\gamma }^{2}}\left( 1+{{e}^{ik}} \right)}{{{\left( -1+{{e}^{ik}} \right)}^{3}}-2\gamma \left( 2-3{{e}^{ik}}+{{e}^{2ik}} \right){{e}^{ik}}+{{\gamma }^{2}}\left( -3+{{e}^{ik}} \right){{e}^{2ik}}} \right|}^{2}}.
\label{eq:double}
\end{align}
\end{subequations}
\end{widetext}

For a single-impurity chain, Fig.~\ref{fig:analytics}(c) 
depicts the reflection (blue) and transmission (red) efficiency for $\gamma=0.5$ (left panel) and $\gamma=2.5$ (right), as a function of the wavenumber $k$.
Both stiffness ratio cases show similar monotonic increase
(decrease) of the reflection (transmission). On the other hand, 
the double-impurity case with $\gamma=0.5$ exhibits a non-monotonic
change of reflection and transmission. In particular, we find that
there exists a reflectionless mode (i.e., the reflection
coefficient is identically equal to zero) at a non-zero wavenumber, 
and complete transmission takes place at that wavenumber,
in a way reminiscent of the Ramsauer-Townsend effect.
However, if we increase the stiffness value from $\gamma=0.5$ to 2.5, these unique features disappear.

To thoroughly analyze the change of reflection and transmission
coefficients, we plot the latter two 
as a function of the stiffness ratio $\gamma$ and wavenumber $k$. 
These are shown in Figs.~\ref{fig:analytics}(e)-(f) for single- and 
double-impurity chains, respectively.
In these figures, the dashed lines indicate regions with $|R|=0$ or $|T|=1$.
The single-impurity chains show monotonic changes of reflection and
transmission for any stiffness ratios as we increase or decrease
the wavenumber from $k=0$. As 
far as the double-impurity chains are concerned, our
analysis shows 
an additional minimum reflection valley and transmission
peak at non-zero wavenumbers if $\gamma \leq 2$.



\subsection{Numerical simulations}\label{sec:simulation}
\begin{figure*}[htbp]
    \centerline{ \includegraphics[width=1.\textwidth]{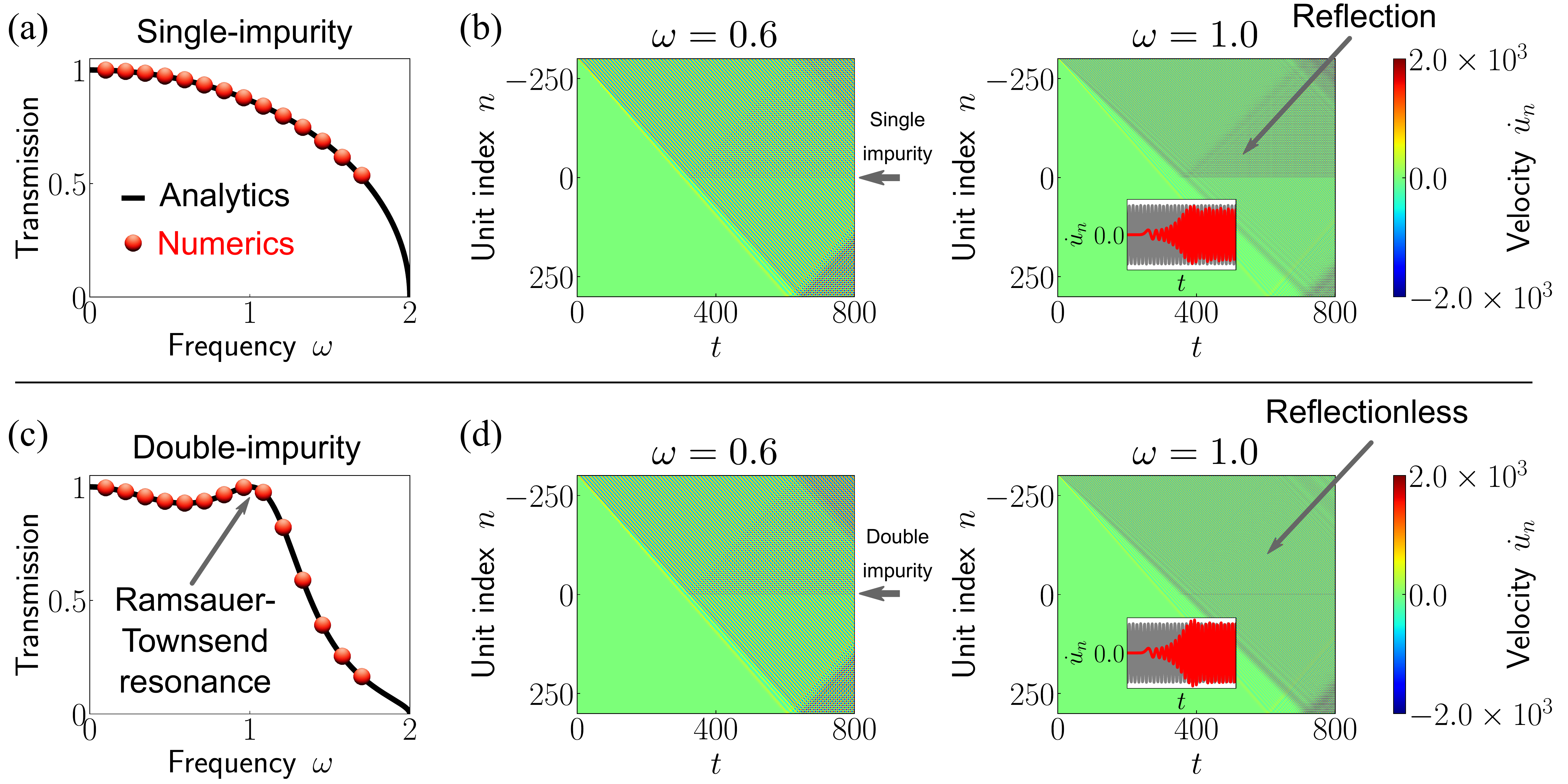}}
    \caption{Analysis of transmission. (a) Comparison between analytics (black line) and numerics (red markers) for a stiffness ratio of $\gamma=0.5$. (b) Spatio-temporal plots of velocity profiles for excitation frequencies of (left) $\omega=0.6$ and (right) $\omega=1.0$, respectively. The inset shows the temporal dependence of the velocity $v_{n}$ for 
    (gray) $n=-269$ and (red) $n=31$. (c-d) Same as the top panels but for the case corresponding to a double-impurity chain. 
    }
    \label{fig:comparison}
\end{figure*}
We now verify the above theoretical considerations by solving
Eq.~\eqref{eq:EqMo_Mono} directly. 
To draw comparisons between the 
theoretical analysis 
and numerical simulations we 
measure the transmission coefficient numerically by analyzing
velocity profiles
of incident and transmitted waves under harmonic 
excitation~\cite{Martinez2016}. For our numerical computations, 
we consider $N=601$ particles ($n \in [-300,300]$) and embed an effective impurity.
The latter is in the form of a compactly supported nodes at the
center of the chain corresponding to $n=0$ for a single-impurity,
and $n=0,1$ for a double-impurity. We apply harmonic excitation 
to the left end of 
the chain ($n=-300$) in the form of a force input 
$F_{ex}=F_0 \sin(\omega t)$ with $F_0=10^{-3}$~N. 
We then 
calculate the transmission coefficient from numerical simulations, 
and analyze the velocity profile of the $n=-269$ particle for
incident waves 
and that of the $n=31$ particle for transmitted waves
(see Appendix B for 
details about the calculation of 
the transmission from 
direct numerical simulations).

In Fig.~\ref{fig:comparison}, we compare theoretical 
results with numerical simulations 
for the stiffness ratio $\gamma=0.5$.
Figure~\ref{fig:comparison}(a) shows  the transmission coefficients
from numerical simulations (red markers) for the single-impurity
case, which demonstrates excellent agreement with the analytical
prediction of Eq.~\eqref{eq:single} (solid black line).
For a single-impurity chain, the reflection coefficient increases 
with frequency (i.e., wavenumber). 
In particular, we observe 
reflected waves with larger amplitude 
for a value of the excitation frequency of $\omega=1.0$, compared with the case with $\omega=0.6$,
as shown in Fig.~\ref{fig:comparison}(b).
In the case of a double-impurity chain, numerical computations capture a
transmission peak at non-zero excitation frequency of about
$\omega=1.0$, denoted by the gray arrow in
Fig.~\ref{fig:comparison}(c). In addition, we observe 
reflected waves for the $\omega=0.6$ case (see the left panel of
Fig.~\ref{fig:comparison}(d)), however, such reflected waves are
unnoticeable if $\omega=1.0$ is applied to a double-impurity
chain (see the right panel of Fig.~\ref{fig:comparison}(d)), which
corresponds to the reflectionless mode of the RT resonance.

\section{Chain with multiple impurities}\label{Sec4}
\begin{figure}[htbp]
    \centerline{ \includegraphics[width=0.5\textwidth]{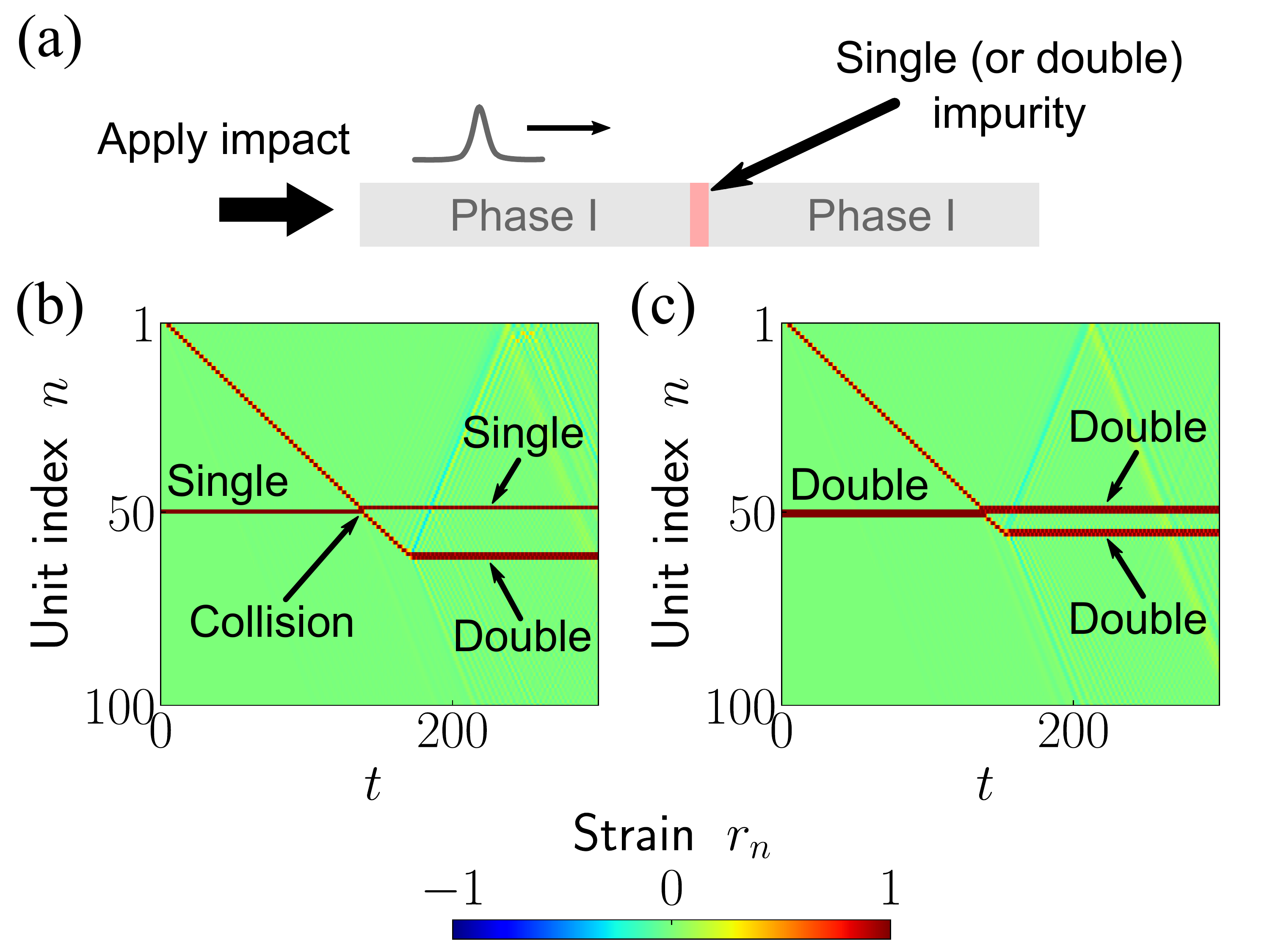}}
    \caption{(a) Schematic illustration of a collision of a 
    propagating wavepacket with an impurity. (b-c) Spatio-temporal plots of strain wave profiles for the (b) single and (c) double impurity cases. We apply the initial velocity ($v_1(t=0)=0.55$) to the first particle to generate the 
    subsonic propagating wavepacket.
    }
    \label{fig:Collision_soliton}
\end{figure}

\begin{figure}[htbp]
    \centerline{ \includegraphics[width=0.5\textwidth]{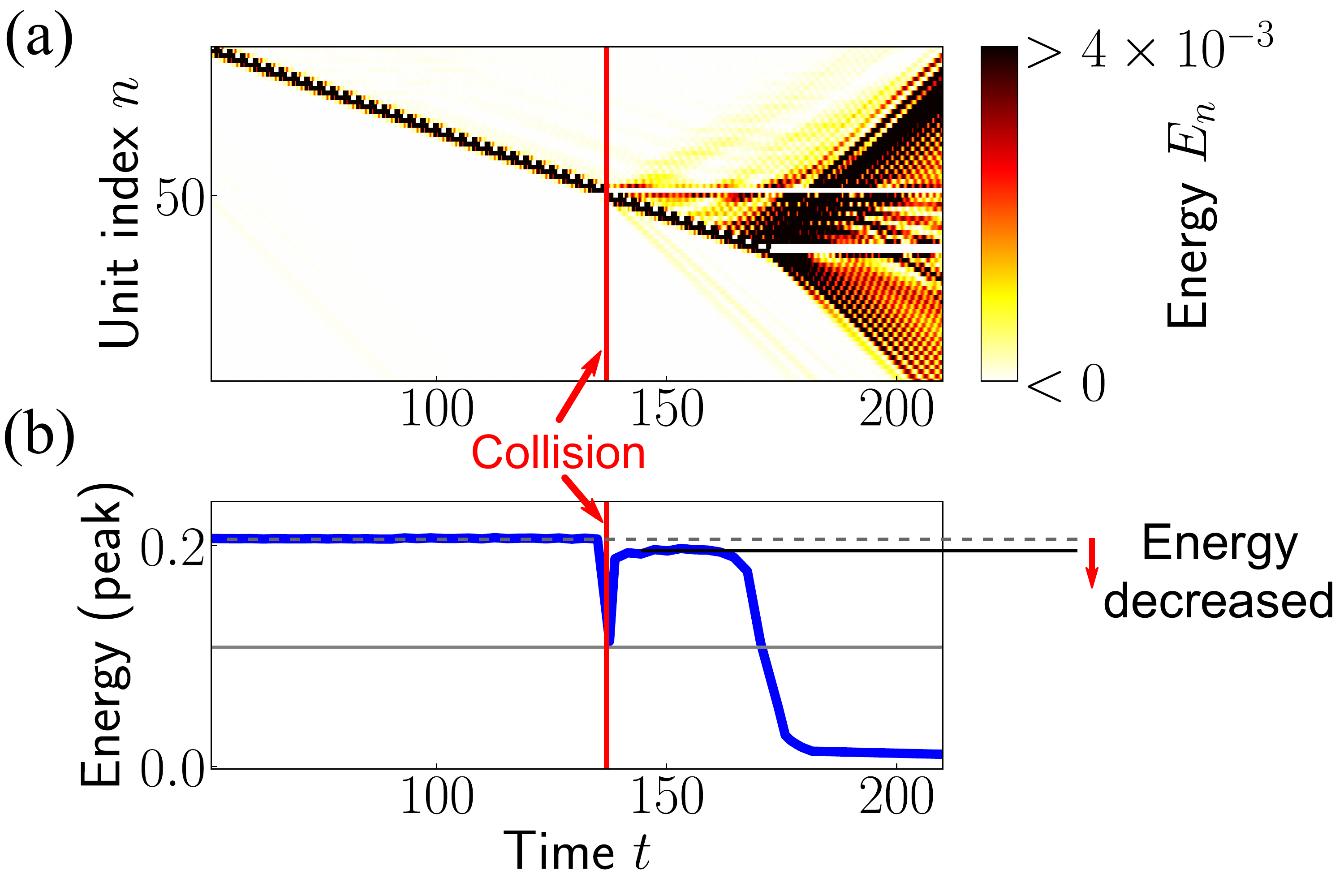}}
    \caption{Collision of a (subsonic) propagating wavepacket with a single impurity. (a) Surface plot of the energy $E_n(t)=\frac{1}{2}m \dot u_n^2(t) + U_n(t)$ where $U_n$ is the bistable potential energy function as shown in Fig.~\ref{fig:piece_wise}(c). The red vertical line indicates the collision point. (b) The peak energy of the propagating wavepacket is shown as a function of time. }
    \label{fig:collision_energy}
\end{figure}

\begin{figure*}[htbp]
    \centerline{ \includegraphics[width=1.\textwidth]{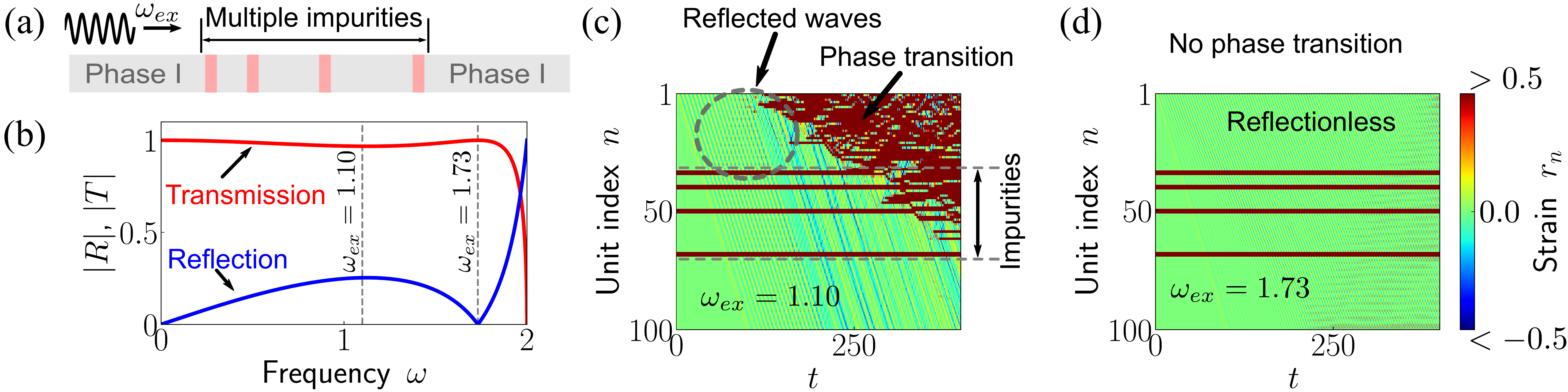}}
    \caption{Scattering by multiple double impurities. (a) The schematic illustration highlights the 
    application of a harmonic excitation (with excitation frequency $\omega_{ex}$) to the first unit ($n=1$) of the chain with multiple double impurities. The stiffness parameters for the host and impurities are $K_{host}=1$ and $K_{imp}=1.5$, respectively. (b) The transmission and reflection coefficients for the double impurity with $\gamma=1.5$ are 
    shown as a function of frequency $\omega$. Surface plots of strain wave propagation are presented 
    for: 
    (c) $\omega_{ex}=1.10$ and (d) $\omega_{ex}=1.73$.}
    \label{fig:Multi_imps}
\end{figure*}

With a firm understanding of scattering by single and double
effective impurities, we consider now the formation of multiple impurities 
in the chain and demonstrate an interesting application of the
reflectionless mode, 
specifically for reconfigurable manipulation of linear wave propagation.
In Sec.~\ref{Sec2}, we observed the formation of a (double) 
impurity due to the trapped energy transported by a 
propagating wavepacket.
In this section, we show that 
the collision of such a 
wavepacket with an impurity can introduce
additional impurities. 

To demonstrate this, we perform numerical
simulations by generating a 
propagating wavepacket in a chain with a single 
or double impurity under impact (see
Fig.~\ref{fig:Collision_soliton}(a)).
The stiffness parameters used in this section are $K_{I}=1$ and $K_{II}=1.5$
with 
an initial velocity of $v_1=0.55$ applied to the first
particle.
This 
generates a 
wavepacket 
as shown 
in 
Fig.~\ref{fig:Collision_soliton}(b) which 
corresponds to 
the single-impurity case.
We find that a 
wavepacket penetrates through a single-impurity
but its propagation is stopped
before reaching the end of
the chain, 
leading to an additional double impurity besides the initial
single impurity. 
Note that the initial impurity is shifted backward by one spatial node.
If 
a 
wavepacket collides with a double impurity,
then an additional
double impurity is formed once the solitary wave passes through 
the impurity region (Fig.~\ref{fig:Collision_soliton}(c)).

We examine such interactions between a subsonic wavepacket and an impurity by considering the energy change before and after the collisions.
In Fig.~\ref{fig:collision_energy}, panel (a) shows the spatio-temporal plot of the total energy $E_n$ for the single-impurity case (corresponding to Fig.~\ref{fig:Collision_soliton}(a)), and we also track the peak energy of the subsonic wavepacket as shown in panel (b). After the collision, some of the energy is reflected (see also the energy decrease of the transmitted wavepacket as shown in panel (b)). However, this reflected energy is not sufficient to generate another propagating wavepacket, and it partly  gets trapped causing the backward
shift of the impurity, which is followed by the trapping of the transmitted wavepacket.
Therefore, our numerical results demonstrate the feasibility of
creating multiple impurities via collisions of 
the propagating wavepacket with
impurities, instead of directly manipulating the phase of the individual units. 

Having confirmed the formation of multiple impurities in our system,
a natural question to ask is whether the reflectionless mode can be observed
in a chain with multiple impurities (see Fig.~\ref{fig:Multi_imps}(a)
for a schematic).
Figure~\ref{fig:Multi_imps}(b) shows
the analytical reflection and transmission coefficients for a
chain with multiple double impurities ($\gamma=1.5$) where 
we identify the reflectionless 
mode at $\omega=1.73$. To examine the scattering between plane waves
and multiple impurities, we consider a chain in which $4$
double impurities 
are embedded. Then, we 
apply two different harmonic excitation inputs to the chain for
comparison: $\omega_{ex}=1.10$ and $1.73$. Here, the amplitude of
excitation force is $F_0=0.14$. Figure~\ref{fig:Multi_imps}(c) shows
the space-time contour plot of strain wave propagation for an excitation
frequency of $\omega_{ex}=1.10$. As predicted, we observe the reflected
waves. Once these waves reach the left end of the chain, the
amplitude of the strain becomes large enough to trigger a phase
transition toward Phase II propagating in the chain due to the amplitude-dependent
nature of our bistable system. On the other hand, if 
$\omega_{ex}=1.73$, the wave passes through the impurity region
without noticeable reflected waves and no particle overcomes the
energy barrier necessary for a phase transition. Once again, our
detailed understanding of the scattering problem enables a characterization
of the associated reflectionless dynamics within the chain.

\section{Conclusions \& Future Directions}\label{conclusion}
In the present work, we investigated the formation of effective impurities 
in a bistable chain arising from the trapping of propagating wavepackets formed upon impact.
We numerically studied the 
amplitude dependence of the impact speed on the behavior of linear and nonlinear wave propagation in the bistable chain whose spring elements exhibit an asymmetric energy landscape.
In particular, we  used numerical simulations
to investigate the propagation of wavepackets
formed upon impact.
Interestingly, our numerical analysis revealed that due to the formation of an oscillatory tail during time propagation, the 
resulting moving wavepacket loses its energy and eventually gets trapped into a lower energy stationary state. 

Based on this emergent localization, we systematically studied the interaction of impurities with 
linear waves and moving wavepackets.
For the linear wave case, we examined the scattering between plane waves and localized 
structures. 
In the system with a single impurity, 
our analytical and numerical results showed that the transmission coefficient monotonically decreases as the harmonic input excitation increases.
On the other hand, the bistable chain with two adjacent units in the lower energy phase exhibits a reflectionless mode due to the analog of the well-known Ramsauer-Townsend effect.

Also, we explored the interaction between a single/double impurity and 
a propagating wavepacket.
The latter can pass through the impurity, instead of being reflected or merged into the existing impurity, however, the wavepacket eventually gets trapped in a low-energy state, which creates an additional impurity in the chain.
Based on this feasibility of adding multiple impurities, together with the reflectionless mode, we further demonstrated that a phase transition front, or a reflectionless
propagation can be ``engineered'' 
in the chain with multiple double impurities, depending on the input harmonic frequency.
Given that the impurities created by 
moving wavepackets in a bistable chain can change the linear/nonlinear wave dynamics, such bistable systems can be useful for controlling a reconfigurable structure in a flexible manner. 
It should be noted in passing that such impurities can be removed from the system by generating the propagation of phase transition fronts~\cite{Yasuda2020} 
(see also Appendix C for numerical demonstrations). This in principle allows reprogramming of the system's linear dynamic response.

This work paves the path for future efforts. From the theoretical perspective, 
the analytical derivation of exact traveling solutions 
in the present setup would be an interesting problem to pursue
by using techniques such as those discussed in Refs.~\cite{VAINCHTEIN2010227,AnnaVPanosK2012}
for a diatomic lattice (see also~\cite{Truskinovsky2014}). 
On the other hand, a systematic existence/bifurcation and stability analysis of traveling waves
will be of particular relevance to the present setting. 
To that end, spectral collocation techniques such as those presented 
in Ref.~\cite{Eilbeck1990} could be employed. 
An especially important aspect of such considerations concerns
the nature of the propagating subsonic wavepackets. While
these spontaneously emerge as a result of the initial impact, 
they also shed energy away only to eventually (and spontaneously) transform themselves
to compactly supported standing excitations. A further understanding
of potential traveling solutions associated with these wavepackets
and of the energetics of the above process would be especially useful
for the further unveiling of the dynamics of such bistable lattices.
Of course, here we have only considered
a piecewise linear problem, while realistic systems could be smooth, nonlinear variants
thereof. It will be especially interesting to explore which of the ramifications of the
present setting (including potential subsonic traveling patterns, compactly supported standing waves
and their respective stability traits and scattering implications) are particular to the
present setting and which generalize, including, potentially, in higher-dimensional
settings.
These directions 
are currently under consideration.

\section*{Acknowledgements}
The authors thank Professor Anna Vainchtein (University of Pittsburgh) for numerous useful discussions. 
HY and JRR gratefully acknowledge the support of ARO grant number W911NF-17–1–0147, DARPA Young Faculty Award number W911NF2010278, and NSF grant number CMMI-2041410. PKP and JRR acknowledge support from Penn's Materials Research Science and Engineering Center (MRSEC) DMR-1720530. This material is based upon work supported by the US National Science Foundation
under Grant No. DMS-1809074 (PGK).\\

\appendix
\renewcommand\thefigure{\thesection A\arabic{figure}}    
\setcounter{figure}{0}

\section*{Appendix A: Amplitude-dependent phase transition}
\begin{figure*}[htbp]
    \centerline{ \includegraphics[width=1.\textwidth]{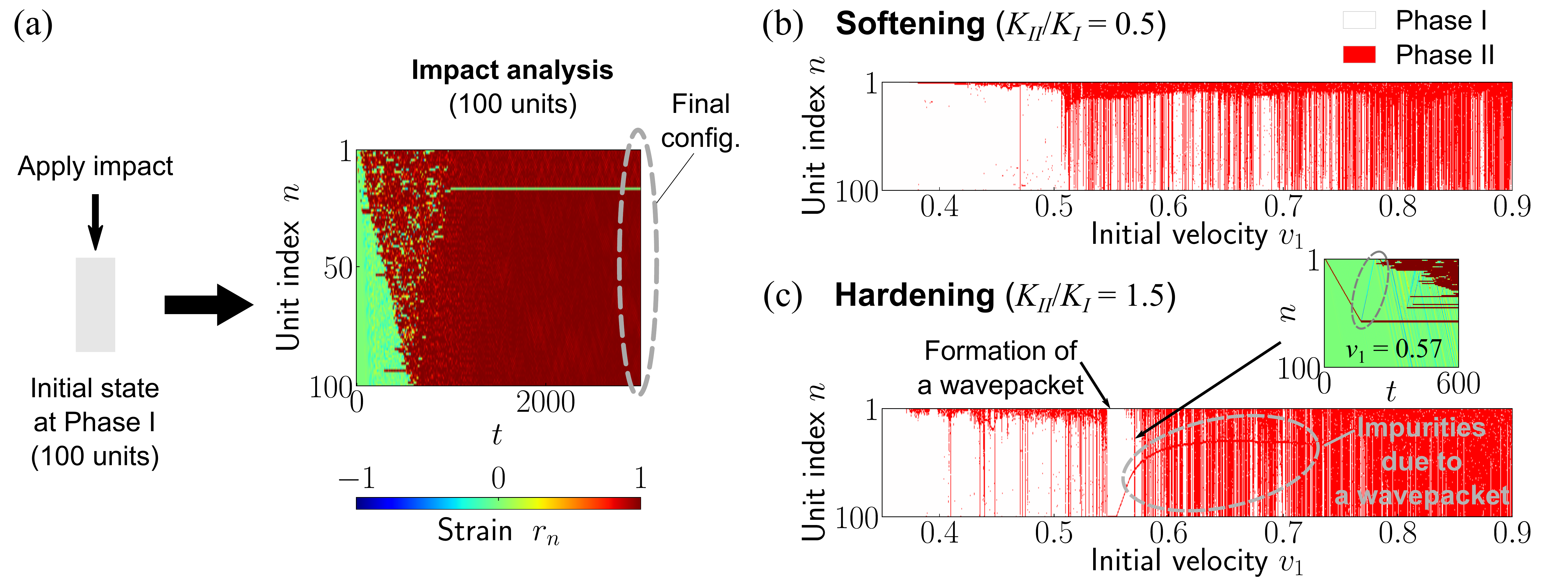}}
    \caption{Amplitude-dependent phase transition behavior. (a) Schematic illustration 
    for examining the steady-state configuration after an impact is applied to the chain. We extract the final configurations of the chain at $t=3000$ and present them as a function of the impact velocity $v_1$ for (b) the softening ($K_{II}/K_{I}=0.5$) and hardening ($K_{II}/K_{I}=1.5$) chains.}
    \label{fig:chain_config}
\end{figure*}

To investigate the impact amplitude-dependent phase transition, we
conduct numerical simulations on a chain with $100$ particles 
by applying various amplitude impacts as shown in
Fig.~\ref{fig:chain_config}(a).
All of the spring elements are initially in Phase I, and 
an initial impact 
is applied to the first particle. 
Numerical simulations are performed for $t \in [0, 3000]$ to allow sufficient time to achieve a 
steady-state 
configuration. We demonstrate the final configuration 
at $t=3000$, 
representing the steady-state configuration after impact, as a function of the amplitude
of impact velocity. 
The corresponding numerical simulation results for 
the softening case ($K_{II}/K_{I}=0.5$) are shown in 
Fig.~\ref{fig:chain_config}(b), whereas 
the those for the hardening case ($K_{II}/K_{I}=1.5$) in
Fig.~\ref{fig:chain_config}(c).
In the figures, the white and red colors indicate the final states 
in Phase I and Phase II, respectively. For the softening case, the phase
transition tends to be localized around the left end of the chain for
smaller amplitude impact, and as we increase the amplitude of the
impact input, the phase transition wave propagates in the chain 
and most of the spring elements move to Phase II states.

For the hardening chain, we obtain more interesting behavior,
specifically the formation of a propagating wavepacket. 
In
Fig.~\ref{fig:chain_config}(c), we find that phase transition occurs
only near the left end of the chain for smaller amplitude impact, which
is similar to the softening case. However, and in the specific amplitude
regime (denoted by the black arrow in the figure), no particle is in Phase II in the
final configuration due to the formation of a (subsonic) wavepacket 
propagating from one end to the other one of the chain.
If we increase the impact amplitude from that regime, 
a (subsonic) wavepacket stops propagating in the middle of the chain, which forms an impurity,
as we discussed in Sec.~\ref{Sec2}.
Note that when the propagation of 
a wavepacket is brought to halt,
waves propagating toward the left end of the chain are generated (see
the red dashed circle in the inset plot of
Fig.~\ref{fig:chain_config}(c)), which sometimes induces phase
transition behavior.

\begin{figure}[htbp]
    \centerline{ \includegraphics[width=0.5\textwidth]{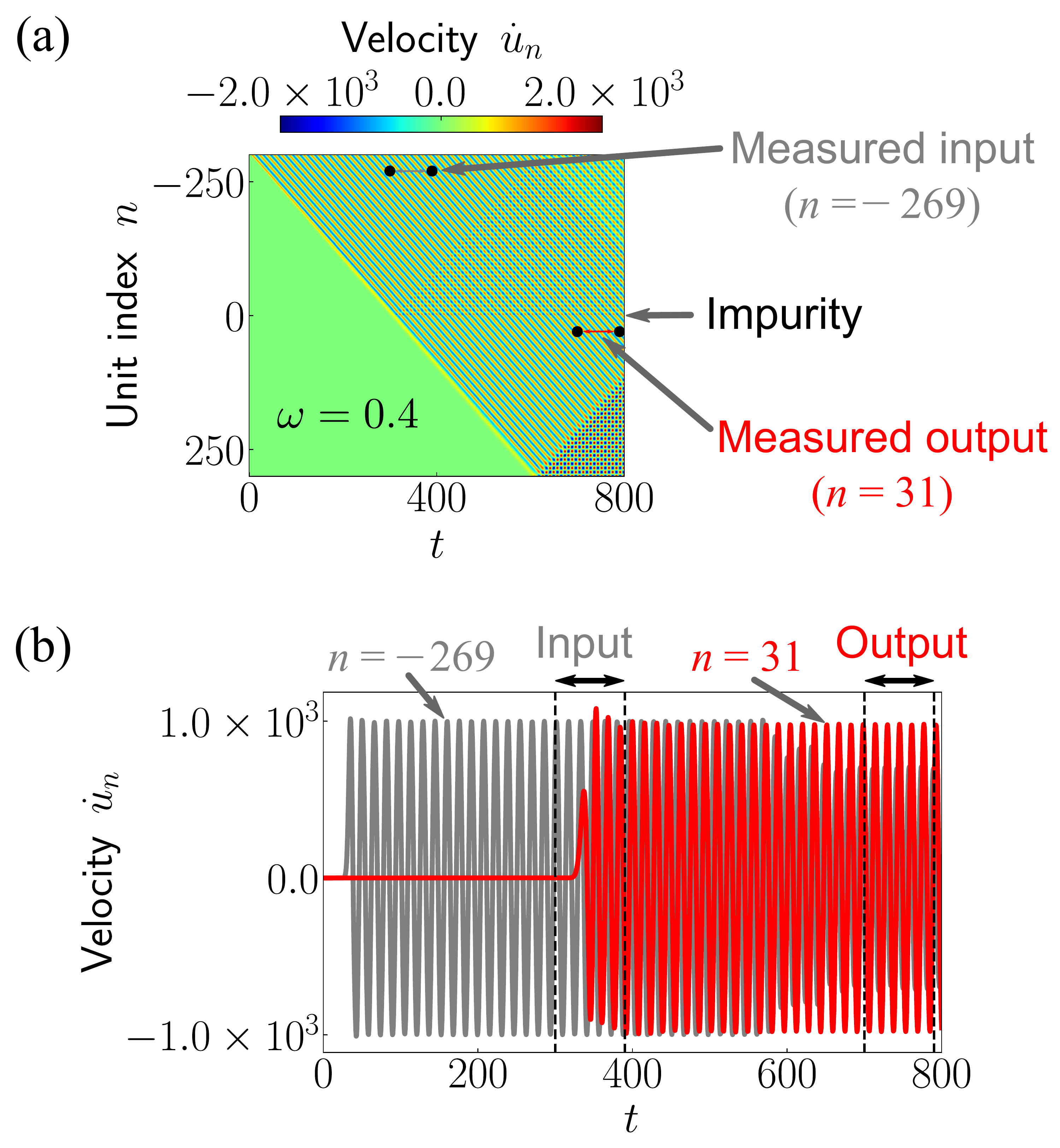}}
    \caption{Numerical simulations to obtain the transmission coefficient. (a) Spatio-temporal plot of velocity. We embed a single impurity (stiffness ratio $\gamma=0.5$) in a chain composed of 601 units, 
    and apply harmonic excitation ($\omega=0.4$) to the chain. (b) The temporal distribution of the velocity profiles $v_{n}$ 
    for: (gray) $n=-269$ and (red) $n=31$.
    }
    \label{fig:measurment}
\end{figure}

\section*{Appendix B: Transmission for various stiffness ratios}
\begin{figure}[htbp]
    \centerline{ \includegraphics[width=0.5\textwidth]{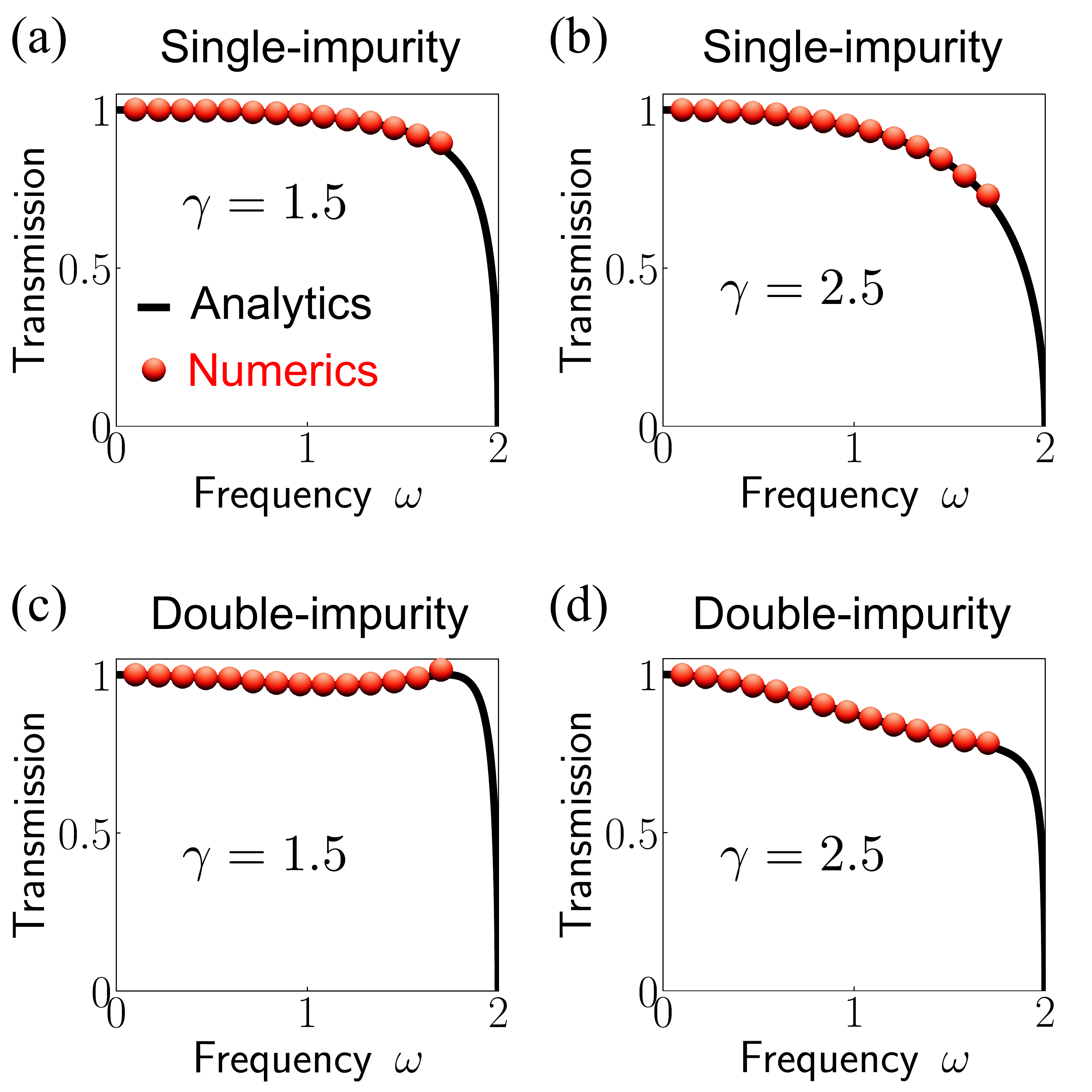}}
    \caption{Analytical transmission compared with numerical simulations for (a,b) single impurity and (c,d) double impurity cases. We calculate transmission coefficients for various stiffness ratios: (left) $\gamma=1.5$ and (right) $\gamma=2.5$.}
    \label{fig:comparison_Appendix}
\end{figure}
Figure~\ref{fig:measurment}(a) shows a spatio-temporal plot of 
particle velocities for the case corresponding to 
$\omega = 0.4$.
To calculate the transmission coefficient from numerical simulations, 
we analyze the velocity profile of the $n=-269$ particle for incident 
waves and that of $n=31$ particle for transmitted waves.
In Fig.~\ref{fig:measurment}(b), we present the temporal distribution
of the velocity profiles $v_{n}$ for the above mentioned values of $n$
where
we calculate the average amplitudes 
from the steady-state region bounded by the vertical dashed lines.
Let $A_{I}$ and $A_T$ be the average amplitudes of the incident 
waves measured at $n=-269$ and transmitted waves at $n=31$. 
Then, we calculate numerically the transmission as $T_{sim}=A_T/A_I$.


Besides the case with stiffness ratio $\gamma=0.5$ discussed in Sec.~\ref{sec3},
we additionally conduct numerical simulations for 
$\gamma=1.5$ and $2.5$. In Fig.~\ref{fig:comparison_Appendix}, 
we show numerical 
results for a chain with (a,b) a
single-impurity, and (c,d) a double-impurity. The 
left (right) two plots are obtained for stiffness ratio of 
$\gamma=1.5$ ($\gamma=2.5$). Similarly to the comparison for
$\gamma=0.5$ in the main text, we confirm excellent agreement 
between numerical simulation results and
analytical transmission coefficients for all four cases.

\section*{Appendix C: Impurity disappearance via propagation of phase boundary}
\begin{figure*}[htbp]
    \centerline{ \includegraphics[width=1.0\textwidth]{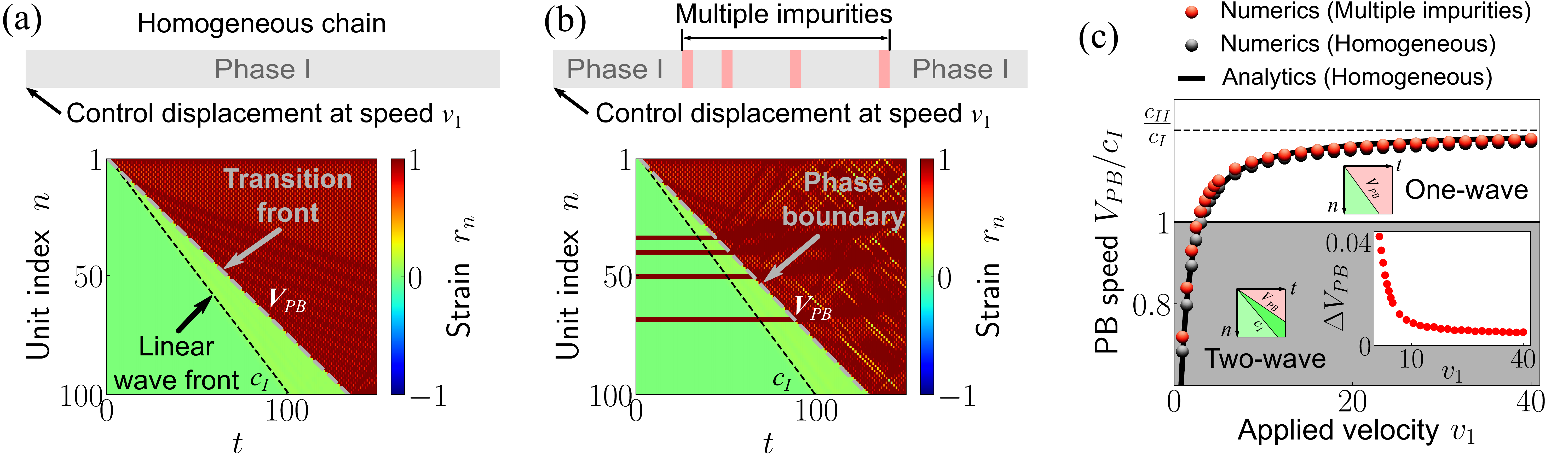}}
    \caption{Effect of multiple double impurities on phase transitions. Spatio-temporal plots for (a) a homogeneous chain (Phase~I) and (b) a chain with double impurities ($K_{imp}=K_{II}=1.5$). We control the displacement of the first unit at constant speed $v_1=1.2$. The gray (black) dashed lines indicate the phase boundary (linear wave front). (c) We plot the speed of the phase boundary ($V_{PB}$) as a function of the applied speed $v_1$. The inset plot shows the difference between the phase boundary speeds for the homogeneous chain without impurities and the chain with multiple impurities, which is defined as $\Delta V_{PB}=V_{PB}^{i}-V_{PB}^{h}$. 
    }
    \label{fig:PhaseTransition}
\end{figure*}

To 
further reveal the nature of a bistable lattice, we
present in this section a way to bring a chain with pre-existing effective impurities
to a homogeneous configuration by utilizing the propagation of phase
boundaries. To achieve this, we control the displacement of the first
particle at constant speed $v_1$. The numerical parameters used in
our analysis are the same as those of Sec.~\ref{Sec4}.
Figure~\ref{fig:PhaseTransition}(a) shows the strain profiles for a
homogeneous chain in which all spring elements are initially in 
Phase~I. We apply 
a constant velocity of $v_1=1.2$ to the first
particle in order to control its displacement. As one can observe 
in other bistable lattices (e.g.,
Refs.~\cite{Slepyan2005,Truskinovsky2006,Deng2020}), phase transitions from Phase~I to Phase~II take place and the phase boundary separating the
phases (denoted by the gray dashed line) moves at constant speed in
the chain. Note that the linear (sonic) waves propagate faster than
the phase boundary (see the black dashed line). By
embedding multiple double impurities in a host homogeneous chain, we
perform the same analysis. In Fig.~\ref{fig:PhaseTransition}(b), 
our numerical results show the propagation of phase boundaries 
in a manner similar to the homogeneous case. These phase
transitions bring all spring elements to Phase~II. In this way, the
bistable chain can
achieve a homogeneous 
configuration (Phase~II) without impurities. Although the overall
strain profiles for the multiple impurities case is similar to that
for the homogeneous case, the propagating phase boundaries
experience phase shifts at the locations of the impurities.


To quantify this phase shifting behavior we calculate the speed 
of the phase boundary propagating in the system by fitting a 
straight line to the phase boundary (see the gray dashed line in
Fig.~\ref{fig:PhaseTransition}(b)). In
Fig.~\ref{fig:PhaseTransition}(c) 
the phase boundary speed $V_{PB}$ is presented as a function of
the applied velocity for displacement control $v_1$.
The gray and red markers indicate the phase boundary speed for a
homogeneous chain and a chain with multiple impurities, respectively.
The solid black line is the analytical phase boundary speed for a
homogeneous chain~\cite{Abeyaratne2006,Zhao2014,Khajehtourian2020}. 
If the applied velocity is small, we find that the phase boundary
speed $v_{PB}$ (obtained from a chain with impurities) 
is higher than that of the homogeneous case 
because the linear waves propagate faster in Phase~II (impurity locations) due to $K_{II}>K_{I}$.
The inset 
panel in Fig.~\ref{fig:PhaseTransition}(c) shows the difference
between the homogeneous and multiple impurities cases,
defined as $\Delta V_{PB}=(V_{PB}^{i} - V_{PB}^{h})/c_{I}$, where
$V_{PB}^{h}$ and $V_{PB}^{i}$ are the phase boundary speeds
calculated from the homogeneous and multiple impurities cases,
respectively.
As 
was already mentioned above, the gap between these two cases is
larger for smaller applied velocity 
and smaller for larger applied velocity, which means that the propagation
of phase boundary is insensitive to multiple impurities in a chain.
This is due to the fact that 
the phase boundary speed at high impact
velocities approaches the sonic wave speed.

\begin{figure}[htbp]
    \centerline{ \includegraphics[width=0.5\textwidth]{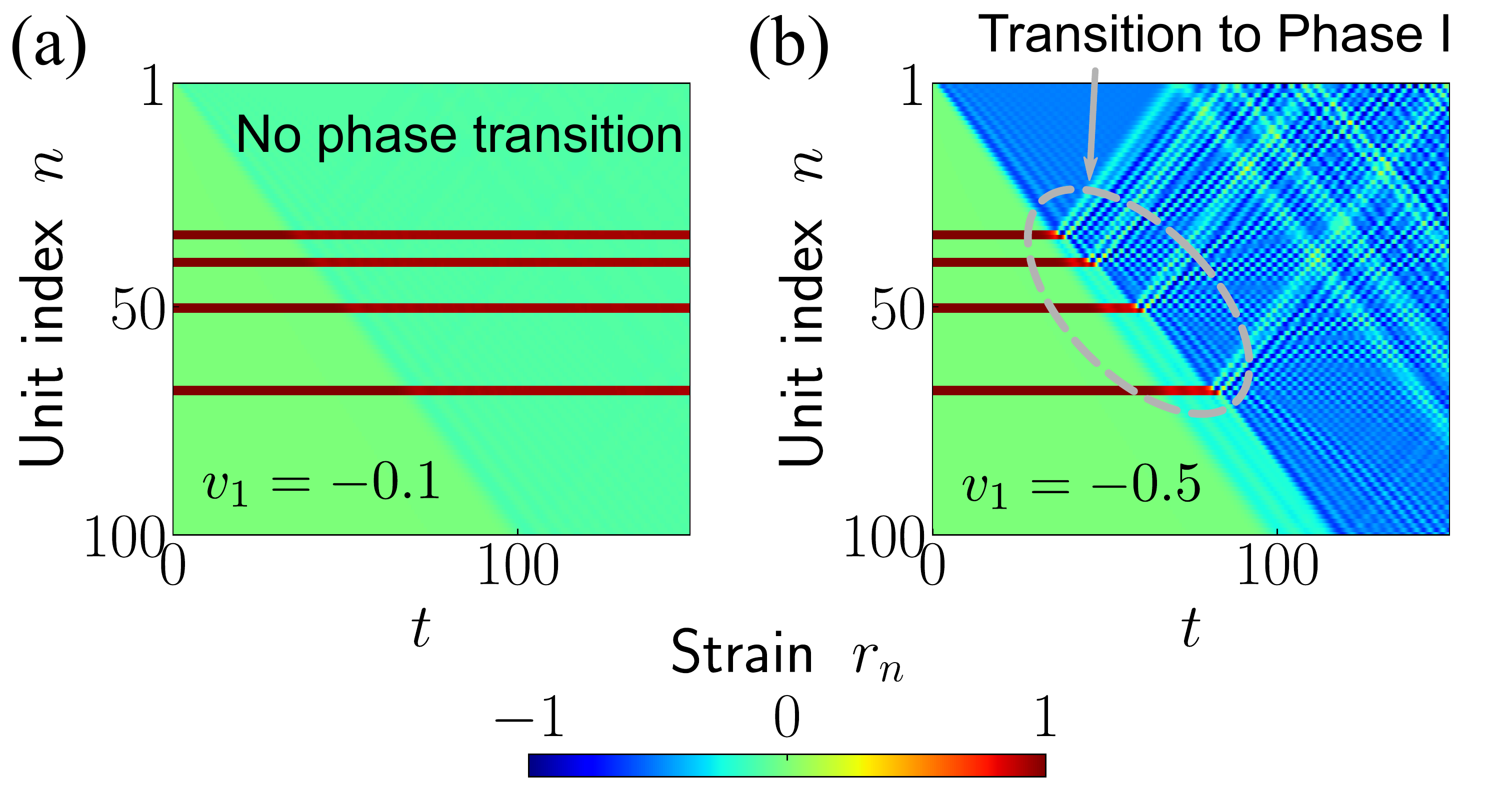}}
    \caption{Initialize a chain with multiple impurities under tension. To apply the tension, we control the displacement of the first particle at constant speed $v_1$. Spatio-temporal plots of the strain wave profiles for (a) $v_1=-0.1$ and (b) $v_1=-0.5$.}
    \label{fig:PhaseTransition2}
\end{figure}

If we control the displacement of the first particle in the opposite
direction it is also possible to bring all spring elements to Phase~I. Figure~\ref{fig:PhaseTransition2} shows the numerical simulation
results for two different applied velocities: (a)~$v_1=-0.1$ and (b)~$v_1=-0.5$. Interestingly, if the applied velocity is not sufficiently 
large, the embedded impurities remain in Phase~II as shown in
Fig.~\ref{fig:PhaseTransition2}(a). However, if $v_1=-0.5$, the 
large-amplitude waves are generated and the impurities are brought
back to Phase~I. Here, we also observe that the waves experience
phase shifts due to impurities.


\bibliography{References}
\break

\end{document}